# Electromagnetically induced transparency in hybrid plasmonic-dielectric system

Bin Tang[1,2], Lei Dai[1], and Chun Jiang[1,*]

[1]*State Key Laboratory of Advanced Optical Communication Systems and Networks, Shanghai Jiao Tong University, Shanghai, 200240, China*

[2] *School of Mathematics & Physics, Changzhou University, Changzhou, 213164, China*

*\*cjiang@sjtu.edu.cn*

**Abstract:** We present theoretical and numerical analysis of a plasmonic-dielectric hybrid system for symmetric and asymmetric coupling between silver cut-wire pairs and silicon grating waveguide with periodic grooves. The results show that both couplings can induce electromagnetically-induced transparency (EIT) analogous to the quantum optical phenomenon. The transmission spectrum shows a single transparency window for the symmetric coupling. The strong normal phase dispersion in the vicinity of this transparent window results in the slow light effect. However, the transmission spectrum appears an additional transparency window for asymmetry coupling due to the double EIT effect, which stems from an asymmetrically coupled resonance (ACR) between the dark and bright modes. More importantly, the excitation of ACR is further associated with remarkable improvement of the group index from less than 40 to more than 2500 corresponding to a high transparent efficiency by comparing with the symmetry coupling. This scheme provides an alternative way to develop the building blocks of systems for plasmonic sensing, all optical switching and slow light applications.

## 1. Introduction

Electromagnetically-induced transparency (EIT) [1] is a quantum optical phenomenon to make an absorptive medium transparent to a resonant probe field owing to the destructive quantum interference between two pathways induced by a coupling field. The importance of EIT stems from the fact it gives rise to greatly enhanced nonlinear susceptibility in the spectral region of induced transparency and is associated with steep dispersion, allowing for many potential applications, such as the transfer of quantum correlations [2], nonlinear optical processes [3], and ultraslow light propagation [4-6]. However, the special experimental conditions for observing this EIT effect in atomic medium hinder its further practical application. This limitation was soon resolved by mimicking the EIT effect in classical oscillator systems [7] that have more merits than in the atomic systems. Recently, various resonant dielectric optical systems including coupled silicon-ring resonators system, photonic crystals, drop-filter cavity-waveguide systems, and a hybridized plasmonic-waveguide system have been proposed and demonstrated to display EIT-like spectral response at room temperature [8-16]. These realizations have further catalyzed an ongoing search for classical systems mimicking EIT. In particular, the plasmonic analogues of EIT in metamaterials, such as dipole antennas [17-23], fish scales metallic patterns [24], split-ring resonators [24-31], trapped-mode patterns [32, 33], and array of metallic nanoparticles [34] which are the most recent and promising additions to the existing array of classical EIT schemes, have been demonstrated theoretically and experimentally. More recently, a method of phase-coupled plasmon-induced transparency has also been presented for achieving the EIT-like in a system of nanoscale plasmonic resonator antennas coupled by means of single-mode silicon waveguide [35].

Thanks to the merging of plasmonics and metamaterials, it opens up a new perspective toward achieving ultimate control of light in the nanoscale dimension. The basic idea of this metamaterial-based EIT-like effect can be understood from two aspects. That is, either arises from normal-mode splitting of a low quality factor (radiative) resonator induced by its coupling with a high quality factor (dark) resonator [17, 18] or from destructive interference between a direct and a resonance-assisted indirect pathways [33]. In fact, the first one is by introducing the EIT-like plasmonic "molecule" consisting of two plasmonic "atoms" supported the radiative state and the dark state respectively. The coupling between the radiative state and dark state could induce the transparent window with strong dispersion and low absorption. The second one is by exciting "trapped mode" resonances [24, 36], which possesses the analogous mechanism and the asymmetric line shape with the Fano effect [37]. View either way, a near-field coupling of the two nanoantennas or resonators is essential in these schemes for observation of the EIT-like spectra. However, in all experimental implementations [24-26, 31, 33] thus far the spectral properties of the dark mode could not be accessed independently since it was exclusively excited by virtue of its coupling to the bright mode. In particular, by breaking the geometrical symmetry of the coupled elements, the exceptional resonance with a very high quality factor can be realized [36]. Such a method provides the additional freedom to manipulate the electromagnetic properties beyond the original resonant responses of metamaterials. In fact, the concept of breaking symmetry has also been applied on other metallic nanostructures, such as asymmetric split-ring resonators [38, 39], nanowire pairs [40, 41] and ring/disk cavity [42], resulting in Fano-type resonances caused by the interaction of narrow dark modes with broad bright modes [37].

As a consequence, in this study we propose a new scheme for generation of the plasmonic EIT in a hybrid plasmonic-dielectric system which is composed of silver cut-wire pairs and silicon grating waveguide with periodic grooves. Our results reveal that the EIT-like effect in our designed system can be achieved, based on the dark-bright coupling mechanisms. In particular, by introducing a structure asymmetry, it is evident to observe a double EIT effect, which stems from an asymmetrically coupled resonance (ACR) between the dark and bright modes [39]. More importantly, the excitation of ACR is further associated with remarkable improvement of the group index from less than 40 to more than 2500 corresponding to a high transparent efficiency. This scheme provides an alternative way to develop the building blocks of systems for applications ranging from plasmonic sensing [43] to all optical switching [44, 45] and slow light applications such as regenerators for optical communication and light storage [46].

## 2. Results and discussion

First, we investigate the case of symmetric coupling in plasmonic-dielectric hybrid system. The proposed structure consists of periodic silver cut-wire pairs and silicon (Si, $n$=3.5) grating waveguide with periodic grooves on one side, which are both embedded in a waveguide material with index of refraction $n$=1.5($SiO_2$). An illustration of the entire structure as well as its profile are shown in Fig.1(a) and (b), respectively. Here, it must be emphatically pointed out that the symmetry and asymmetry coupling we will talk about later are relative to the $y$ direction. In our analysis the resonant response of plasmonic-dielectric system was simulated by using the finite-difference time-domain (FDTD) method [47]. We describe the complex optical constants of metal Ag taking from experimental data [48]. A plane electromagnetic wave is incident on the unit cell along the normal (-$y$) direction, and transmitted power is monitored using a power monitor placed behind the unit cell. Periodic boundary conditions are employed in the simulations along the $x$-axes of the single unit cell to mimic a periodic arrangement. The perfectly matched layers are implemented on the top and bottom of the computation domain to eliminate undesired reflections from the domain boundary. Once the electromagnetic fields are solved, the reflectance $R$ and transmittance $T$, the ratios of reflected and transmitted powers, respectively, to the incident one are determined by the fields at the far boundary. According to the conservation of energy, the absorbance $A$ in the system is given as $A = 1−R−T$.

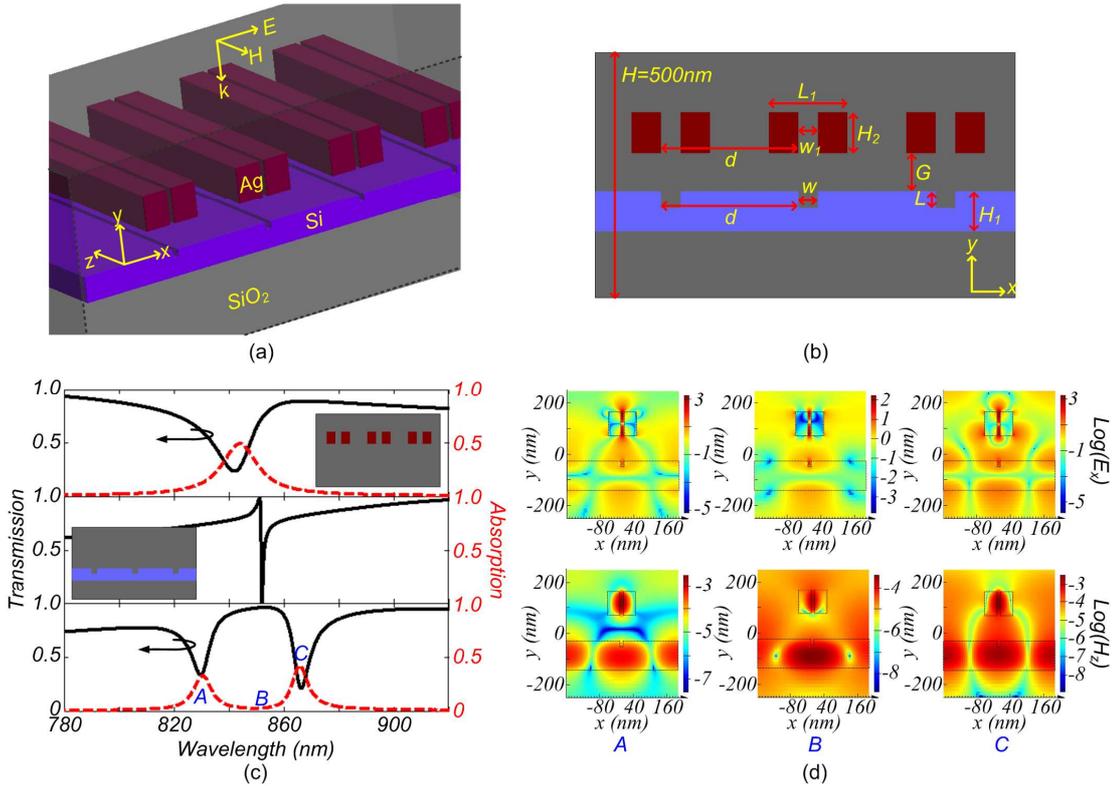

Fig.1 (Color online) (a) Three-dimensional and (b) two-dimensional views of the proposed plasmonic-dielectric system. The geometric parameters are $H_1$=120nm, $H_2$=100nm, $L$=20nm, $L_1$=190nm, $d$=400nm, and $w$=$w_1$=10nm. The vertical distance between the upper silver cut-wire pair and the lower silicon grating is $G$=100nm and the entire thickness of the dielectric slab waveguide is denoted by $H$. (c) The transmission and absorption spectra of a sole silver cut-wire pair(top panel), a sole Si grating (middle panel), and an entire hybrid plasmonic-dielectric system (bottom panel). The insets show the illustration of a sole silver cut-wire pairs and a sole Si grating embedded in $SiO_2$ dielectric slab, respectively. (d) The distribution of the logarithm of electric field and magnetic field at point A, B and C in Fig.1(c)

Based on the electromagnetic field theory, the incoming electromagnetic wave with electric field polarization in the $x$ direction, can excite localized surface plasmon polarations (SPPs) in each metal cut-wire pair, leading to a broadband strong absorption peak in the spectrum as shown in Fig.1(c) (top panel). Actually, a unit cell of metal cut-wire pair constructs a metal-dielectric-metal (MDM) plasmonic resonance cavity which holds an excited cavity mode as a result of the resonant excitation of the fundamental SPPs mode. Meanwhile, truncation of such a MDM waveguide results in strong reflection from the terminations and gives rise to resonant cavity behavior. However, the mode supported by the silicon grating waveguide has bigger quality factor than the case of the case of periodically arranged MDM cavities. Therefore, the upper silver cut-wire pair serves as a bright mode or radiative state in EIT-like plasmonic system. The resonance frequency of the metal cut-wire pair can be readily tuned by varying its spatial dimension and the permittivity of the surrounding media. The lower silicon grating waveguide structure consists of periodic grooves arrays with nominal width $a$ and separated distance $d$ and acts as a dark mode in our classical analog of EIT since it couples weakly with external light and possesses a very long lifetime (narrow linewidth) as shown in Fig.1 (c). In fact, in developing an analogy to EIT a significant difference in the quality factors (or linewidth) of the two resonances involved is required [27, 28].

When the periodic plasmonic resonance cavities are placed on top of the Si grating waveguide, the resulting system depicted in Fig.1 (a) and (b) exhibit the transmittance and absorbance curves of Fig.1(c) (bottom panel) in single period. We observe a broad spectral hole in the absorbance curve (maximum of the corresponding transmittance curve) centered at the middle of the absorbance curve. The emergence of this transparency window is based on the splitting of absorption peak of radiative state. To reveal the origin of the symmetrically coupled resonance (SCR), we simulate the distribution of the $x$-component electric field and the $z$-component magnetic field at the wavelengths of the two transmission dips and at the peak wavelength of the transparency window in Fig.1(d), respectively. (For visuality, we give the distribution of the logarithm fields, similarly hereinafter). From the electric field $E_x$ distribution, we can observe that along the periodic direction there is an excited propagating surface mode that is concentrated on the Si grating waveguide interface, and in the MDM cavity there exist localized SPPs and an excited SPPs-induced cavity mode, respectively. Note that the whole field is the sum of bright mode excited by incident light and the dark mode excited by the bright mode. The electromagnetic field is coupled back and forth between the bright and dark mode, leading to a destructive interference. Thus, based on the mechanism of Fano effect, slow light effect can be achieved at the resonance in the hybrid plasmonic-dielectric system. In particular, at the point A as shown in Fig.1 (c), the propagation dip results mainly from the coupling between the resonant cavities modes and the surface propagating waveguide modes. In contrast, at the higher wavelength propagation dip, that is, at the point C, the propagation dip results mainly from the coupling between the surface propagating waveguide modes and the localized SPPs modes on both sides of the MDM. Surprisingly, the lower propagation dip position almost keeps the same with before, while the coupling phenomena takes place a special change at the higher propagation dip when the structure symmetry is broken. This property will be further discussed later.

The coupling between the dark and the radiative resonators is mainly the near field interaction. In contrast to atomic EIT, where the coupling between the energy states is realized by a pump beam, the EIT-like feature in the plasmonic-dielectric hybrid system arising from the coupling between the radiative modes and the dark modes implies the possibility for the coupling strength to be tuned by changing the distance $G$. In Fig.2(a) and (b), we show the simulated transmission and absorption spectra for this structure. For all separations, the coupling between the radiative and dark modes leads to a broad absorption dip, which confirms the EIT-like destructive interference between the two pathways: direct excitation of the radiative mode from the incident wave and excitation of the dark mode (by the radiative mode) coupling back to the radiative mode. At small separation between the dark Si grating waveguide and the radiative resonators, i.e., strong coupling, the splitting of the two resonances is large. Simultaneously, a strong dip in the absorption spectrum appears. With decreasing coupling between them, the dip in the absorption spectrum becomes narrow. Meanwhile, the corresponding transmission phase experiences a strong change and becomes more and more obvious for the distance $G$ increasing from 100 nm to 160 nm, which can be seen in Fig.2(c). What is more, for a fixed distance $G$, we can observe from the phase spectrum that there are two frequency segments corresponding to the normal phase dispersion and anomalous phase dispersion around in the transparency window. This strongly normal phase change results in a group index more than 30 at the transmission peak on the analogy of the EIT in atom system. It corresponds to an increased traversing time of light through the entire structure. Fig.2(d) gives the group index ($n_g$) versus wavelength for

different coupling separations according to the following formula:

$$n_g = \frac{c}{v_g} = \frac{c}{H}\tau_g = -\frac{c}{H}\cdot\frac{d\varphi(\omega)}{d\omega},$$

where $c$ is the speed of light in vacuum, $v_g$ is the group velocity in the media, $\tau_g$ is the delay time, and the phase $\varphi(\omega)$ is the function of the angular frequency $\omega$. $H$ is the thickness of the plasmonic-dielectric waveguide system.

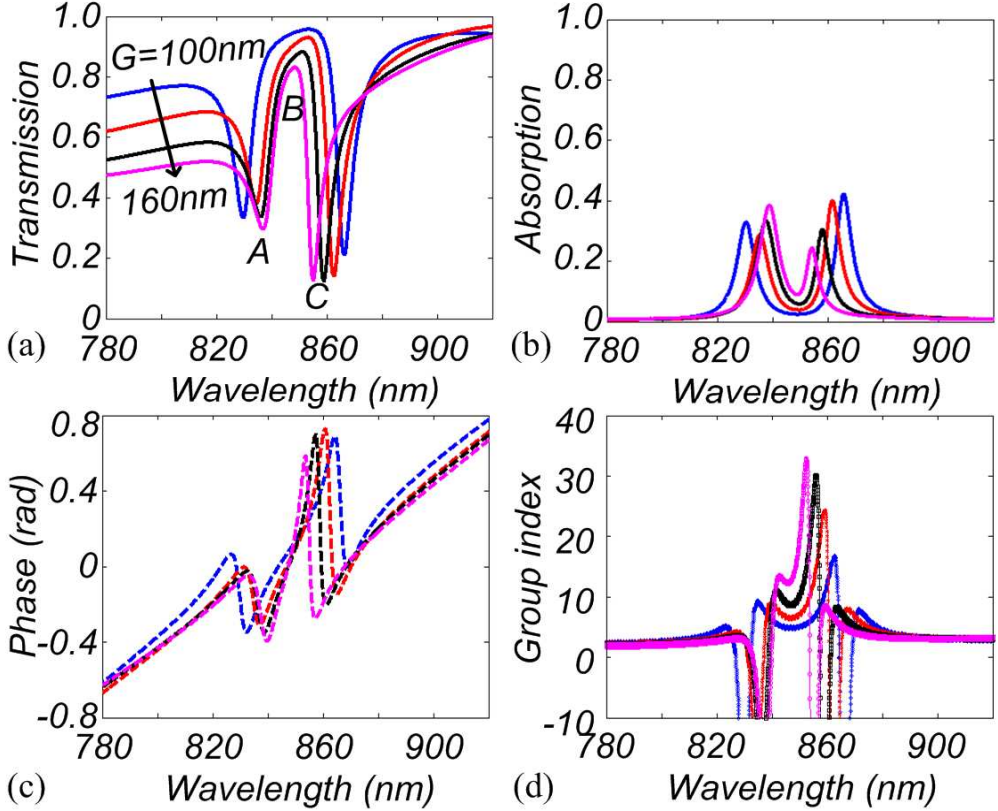

Fig.2 (Color online) (a) Transmission, and (b) absorption spectra for the hybrid plasmonic-dielectric system for different separations. (c) The dispersion of the phase $\varphi(\omega)$ and (d) the group index for the EIT system shown in Fig.1(a).

Unfortunately, the group index obtained from above scheme is very low although it can slow down the speed of light to some extent. To further improve the group index, we make a minor modification of the proposed structure by introducing a structure asymmetry, that is, breaking symmetry, as shown in Fig.3(a). The upper periodic plasmonic resonance cavities waveguide are displaced a slight distance relative to the lower Si grating waveguide in the *x*-direction. In fact, the misalignment between the two waveguides is virtually unavoidable in fabrication; the influence of weakly asymmetric structural elements has to be considered. The lateral displacement of the resonance cavities with respect to the symmetry axis is defined as $G_1$. The other parameters always keep the same as those in Fig.1(a) and Fig.1(b). For comparisons, we plot the transmission and absorption spectra of a plane wave passing through the symmetric and asymmetric structure, respectively. Obviously, From Fig.3(b), we can see that there exists only one transparency window between two transmission dips as discussed before for the case of symmetry coupling ($G_1=0$ nm). However, the results have an evident change for the case of asymmetry one, say $G_1=20$ nm, in which the transmission spectrum appears two transparency windows, that is, double EIT effect [14, 15]. Interestingly, at the higher wavelength transmission dip, there appears another propagation peak, that is, a new appeared transparent window which is based on the splitting of corresponding absorption peak. Meanwhile, there also emerges two propagation dips corresponding to two new appeared absorption peaks on both sides of the new transparent window. Comparing the two transparency windows in asymmetry coupling, the new appeared one becomes much narrower than the former, and the transmission efficiency almost keeps the same and attains more than 85%. Fig.3(c) shows the corresponding phase changes for the two different coupling cases. From the transmission and phase spectra it can be recognized that strong dispersion, leading to a huge group index up to 176, occurs in the new transparency window. This indicates that a light pulse with a center frequency situated in the transparency window will be considerably slowed down upon traversing the plasmonic-dielectric system. To gain a deep insight into the underlying physics of the electromagnetic response of the plasmonic-dielectric system and to understand the double EIT-like effect, we also compute the distribution of the logarithm of the *x*-component electric field and the *z*-component magnetic field at the new appeared transmission dips and transmission peak as shown in Fig.3(d). The structure asymmetry results in the asymmetric coupling resonances at the new appeared propagation dips, that is, point *D* and *F*, respectively. It is evident that the interference of these two resonances, which both lead individually to transmission dips, result in an extremely narrow transparency window in connection with the double EIT-like effect.

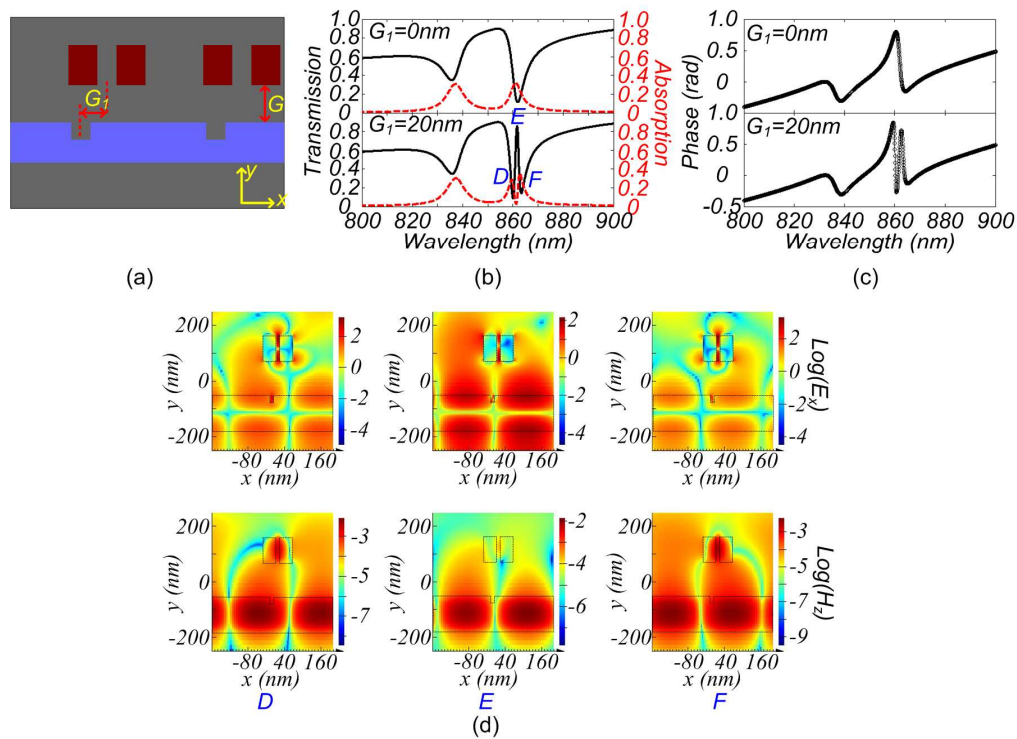

Fig.3 (Color online) (a) Two-dimensional view of the plasmonic-dielectric system with breaking symmetry. The displacement with respect to the symmetry axis is defined as $G_1$. (b) Transmission and absorption spectra for symmetry and asymmetry couplings, and (c) corresponding transmission phase changes with wavelength for $G_1$=0 nm and $G_1$=20 nm, respectively. (d) The distribution of the logarithm of electric field and magnetic field at point *D*, *E* and *F* in Fig.3(b)

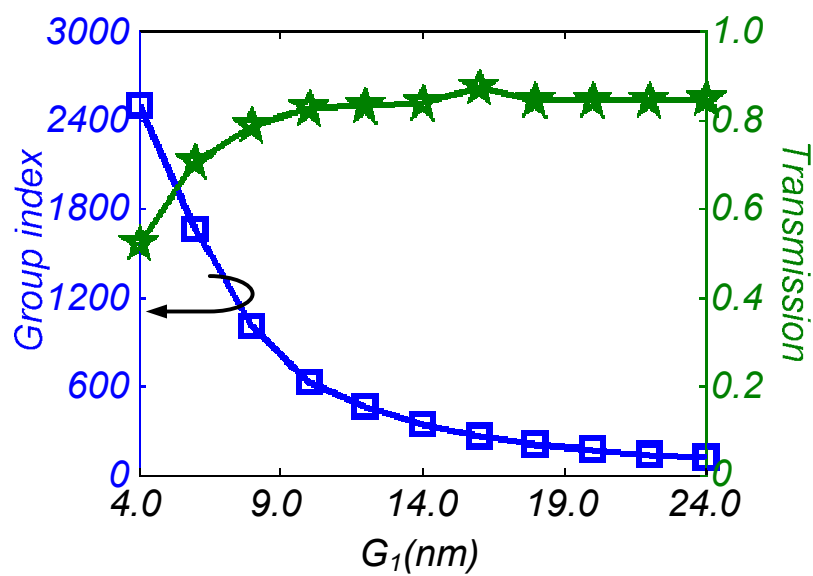

Fig.4 (Color online) The group index change for different degree of asymmetry (blue curve) and its corresponding maximum transmission efficiency (green curve).

From above results, the new appeared sharp transmission peak is associated with a dramatic change in the transmission phase (refractive index), consequently, extremely slow group velocity of light traversing the system can be achieved. Fig.4 shows a more detailed analysis of the group index versus degree of asymmetry (blue curve) and its corresponding maximum transmission efficiency (green curve).The result indicates that the group index and the maximum transmission efficiency are closely related with the degree of asymmetry. In particular, we note that the hybrid plasmonic-dielectric system support the slow light with group index 2594 (of the order of $10^3$) at the lower degree of asymmetry as seen in Fig.4, say $G_1$=4 nm; the corresponding transmission efficiency are also considerable and up to 53%. In contrast, when the degree of asymmetry $G_1$=20 nm, the compound system can achieve slow light with group index nearly 176 and transmission efficiency more than 85%. Therefore, we can conclude that there is a mutual restraint between the transmission efficiency and group index. That is, there is an approximately opposite developing trend between the transmission and the group index when the displacement $G_1$ changes.

## 3. Conclusion

In conclusion, we introduce the coupling of plasmonic resonances in silver cut-wire pairs and silicon grating waveguide with periodic grooves on one side, especially in the asymmetric one that supports an extraordinary electromagnetic response referred to as asymmetrically coupled resonance (ACR), which can lead to a double EIT effect. Similar to the EIT effect in quantum systems, the hybrid plasmonic-dielectric hybrid system exhibits strong phase dispersions and a significantly reduced group velocity; and it can support the slow light with group index more than 2500 and high transmission efficiency in excited transparent window. By artificially mimicking the bright and dark modes, we observe that the ACR response is excited in case of strong coupling between a narrow dark mode with a broad bright mode, and this ACR can be modulated by varying the spacing of two waveguide constituents. This scheme provides an alternative way to develop the building blocks of systems for plasmonic sensing, all optical switching and slow light applications.

## Acknowledgements

This work was supported by National Natural Science Foundation of China (Grant No. 60672017) and sponsored by Shanghai Pujiang Program.